# Cell-in-cell structures are involved in the competition between cells in breast cancer

S. Sajedeh Mousavi[1], Sara Razi[2]

Department of veterinary medicine, Garmsar branch, Islamic Azad university, Garmsar-Iran

Department of Biology, Science and Research branch, Islamic Azad University, Tehran-Iran

Shahid Beheshti University of medical sciences, Tehran, Iran

Sajedeh.mousavi@yahoo.com

**Abstract**

Breast cancer is the most common cancer in women worldwide, and discovering the biomarkers of this disease became so vital nowadays and Cell in Cell structure could be one of them, and it may be used as an available proxy for tumour malignancy. (CICs) are unusual in that keep morphologically healthy cells within another cell. They are found in various human cancers and result from active cell-cell interaction, and it has different kinds. In this study, we analyzed the microarray data from GEO (GSE103865) to genetically evaluate CICs' incidence in samples obtained from breast cancer patients to understand the relationship between the rate of CIC and the prognosis of breast cancer. The preprocessing was performed using R software. The DAVID website was used to analyze gene ontology (GO) and Gene and Genome (KEGG) pathway. The protein-protein interactions (PPIs) of the obtained DEGs were assessed using the STRING website, and hub modules in Cytoscape and cytoHubba were screened. According to the results from analyzing the 20 hub genes, we understood that overexpression of our Top genes is effective in focal adhesion, ECM-receptor interaction, platelet activation and PI3K-Akt signalling pathway, which shows that changes in these pathways could be the reason the overexpression of CICs in breast cancer. These data and research by many others have uncovered various genes involved in CIC formation and have started to give us an idea of why they are formed and how they could contribute to breast cancer.

**Introduction**

Breast cancer tissues are highly heterogeneous, which is crucial for clinical diagnosis and treatment. Cannibalism of cells is a rare pathological condition found in a variety of human tumours, including breast cancer (Huang, Chen et al. 2015). Cannibalistic cells often form cell in-cell (CIC) structures characterized by one cell enclosing another, impacting patient prognosis (Zhang, Niu et al. 2019). CIC structures have been discovered in both human and animal tumours (Ferreira, Soares et al. 2015), implying that CIC is a common malignant phenomenon in all animals.

Cell-in-cell structures (CICs) are unusual in that those who keep morphologically healthy cells within another cell. They are found in various human cancers and result from active cell-cell interaction (Huang, Chen et al. 2015). CICs may form homotypically between tumour cells or heterotypically between different types of cells, such as tumour cells and immune cells, due to the complex cellular composition of cancer tissues (Huang, Chen et al. 2015, Fais and Overholtzer 2018). CIC formation usually results in the death of internalized cells in a non-autonomous manner, with the outer cells killing the inner cells (Overholtzer, Mailleux et al. 2007). After apoptosis (type I), necrosis (type

II), and autophagy-dependent cell death (type III), CIC-mediated cell death was proposed as the type IV cell death (Martins, Raza et al. 2017). CICs seem to have opposing functions (suppressive and proliferative) in human tumours, which are likely influenced by various factors such as CIC type, tumour type, stage, and so on (Huang, Chen et al. 2015, Fais and Overholtzer 2018). The development of a specific kind of CIC results from the integration of multiple signals (Ruan, Wang et al. 2018, Mackay and Muller 2019), and it may be used as a functional proxy for tumour malignancy (Sharma and Dey 2011, Jose, Mane et al. 2014). According to accumulating evidence, CICs were consistently linked to tumour grades (Abodief, Dey et al. 2006, Kim, Choi et al. 2019).

In this study, we analyzed the microarray data from GEO (GSE103865) to genetically evaluate CICs' incidence in samples obtained from breast cancer patients and also by examining the superior upregulated genes and their cell pathways, understanding the relationship between the rate of CIC and the prognosis of breast cancer.

**Material and methods**

*Dataset*

The microarray dataset used in this essay was generated by Sun Q et al. Based on the GPL570 Affymetrix Human Genome U133 Plus 2.0 Array platform and is available in NCBI's Gene Expression Omnibus (GEO) repository, with accession number GSE103865. Breast tumour tissue samples were collected from patients at the Military 307 Hospital, Beijing, and selected according to low and high CICs formation rate for RNA extraction and hybridization on Affymetrix microarrays. We had attained 7 samples for L-CICs and 7 samples for H-CICs.

*Preprocessing and Selection of Differentially Expressed Genes (DEGs)*

We performed some Quality Control (QC) checks involving sample-level plots. These checks comprise a Principal Component Analysis (PCA) plot and boxplot. PCA helps us distinguish samples using expression variations and determine whether the L-CICs and H-CICs samples are differentiable after normalization (Figure 1). The gene expression ranges had been transformed to a logarithmic base 2 scale. The preprocessing was performed using R (version 4.0.3), and we used different packages such as genefilter (version 1.72.1) for filtering genes from though put experiments, affy (version 1.68.0), Affytools (1.62.0), AffyPLM (1.66.0) for fitting prob-levels models and limma (3.46.0) for evaluating DEGs between L-CICs and H-CICs samples. Also, the hgu133plus2.db annotation package was used to convert probe identifiers into gene symbols. The top 200 genes were selected to be submitted to the STRING database. The cut-off criteria for defining DEGs were adjusted P<0.0001 and |logFC| <=1.

*PPI Undirected Unweighted Network Reconstruction of DEGs*

Protein-protein interactions (PPIs) for overlapping DEGs were obtained from the online Search Tool for the Retrieval of Interacting Genes (STRING, http://string-db.org). These interactions were input into Cytoscape software (http://www.cytoscape.org/) to create a PPI network of hub genes. This network was analyzed by the Cytoscape MCODE plugin (degree cutoff = 2, max. Depth = 100, k-core = 2, and node score cutoff = 0.2.). We used the plugin cytoHubba to select the top 20 upregulated hub genes from the PPI network.

To consider our top genes, Biological process, Cellular component, Clinical phenotype, Molecular function, and Site of expression, we used Funrich (version 3.1.3) and the David website.

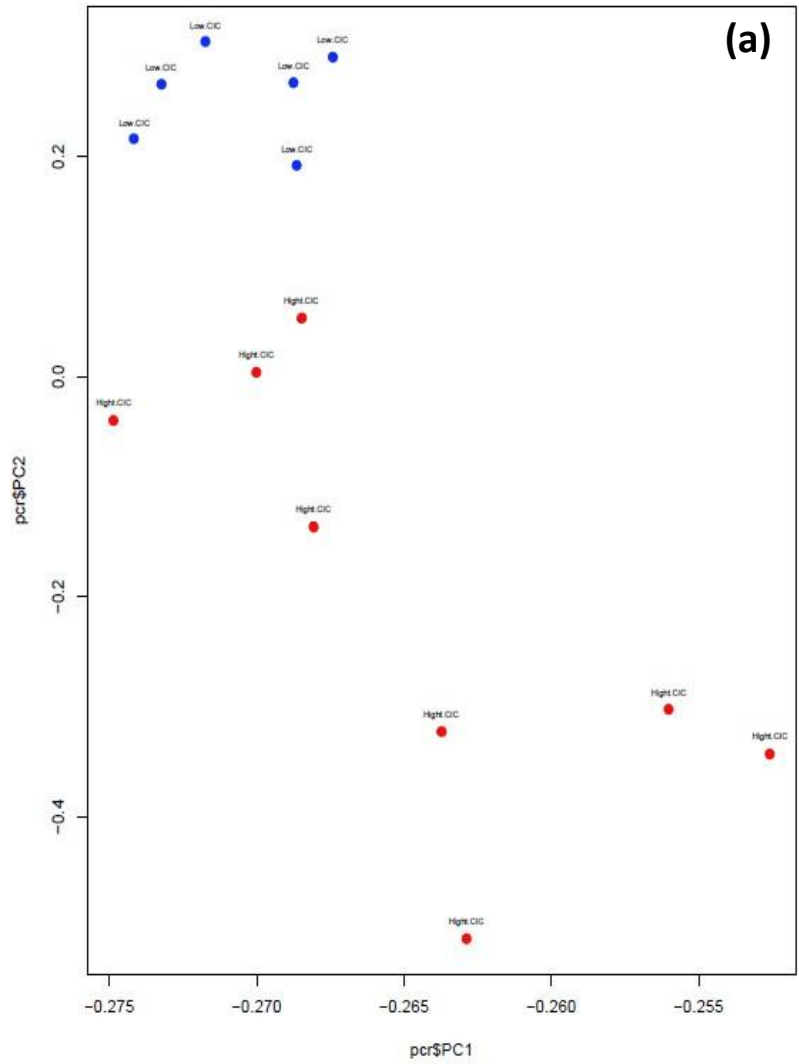

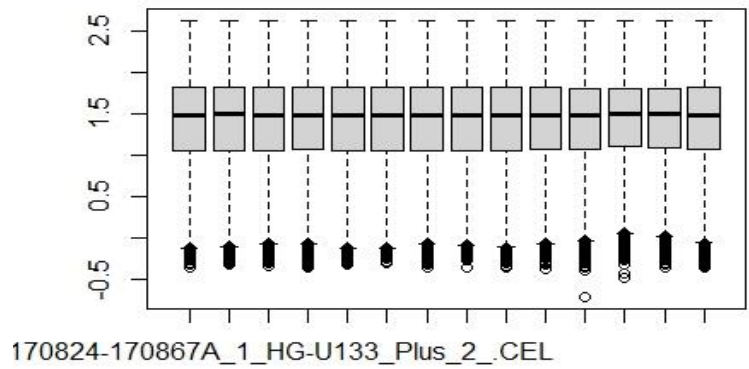

*Figure 1: (a): PCA plot which used to distinguish samples using expression variations and determine whether the Low-CICs and High-CICs samples are differentiable after normalization (blue dots: Low CICs, Red dots: High CICs) / (b): Boxplot for checking the normalization of our samples.*

**Results**

*DEGs gene ontology and KEGG pathway analysis*

To gain further insights into DEGs' biological function, a total of DEGs consisted of 124 DEGs were submitted into DAVID software for GO analysis and KEGG pathway analysis. In biological process (BP) term, the result indicated that the DEGs were mainly enriched in Collagen catabolic process, extracellular matrix organization, extracellular structure organization, and multicellular organism catabolic process. (Figure 2.a). Regarding the cellular component (CC) term, the DEGs were mainly involved in intracellular activations (Figure 2.b). About molecular function (MF), the DEGs were primarily associated with extracellular matrix structural constituent, collagen binding, platelet-derived growth factor binding, and RAGE receptor binding (Figure 2.c). The DEGs were mostly linked to ECM-receptor interaction, Protein digestion and absorption, Focal adhesion, Amoebiasis, Platelet activation, Renin secretion, Arrhythmogenic right ventricular cardiomyopathy (ARVC), PI3K-Akt signaling pathway, Glutamatergic synapse and Long-term depression according to the KEGG pathway research. (Figure 2.d).

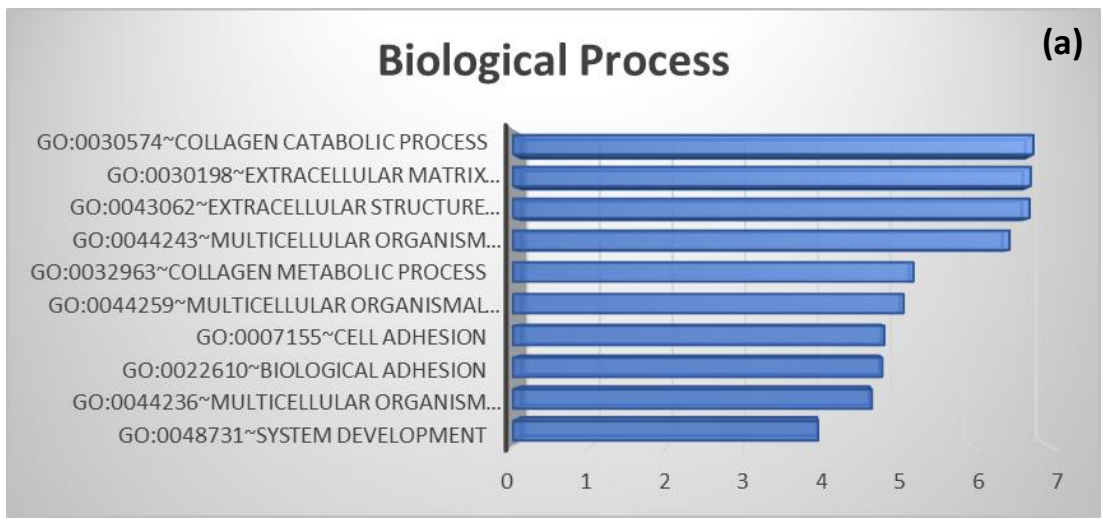

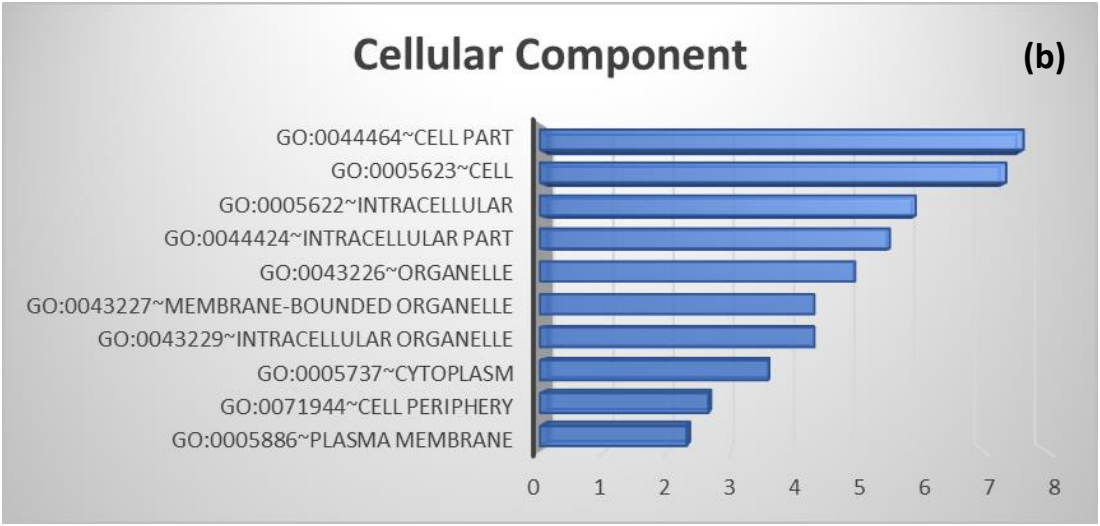
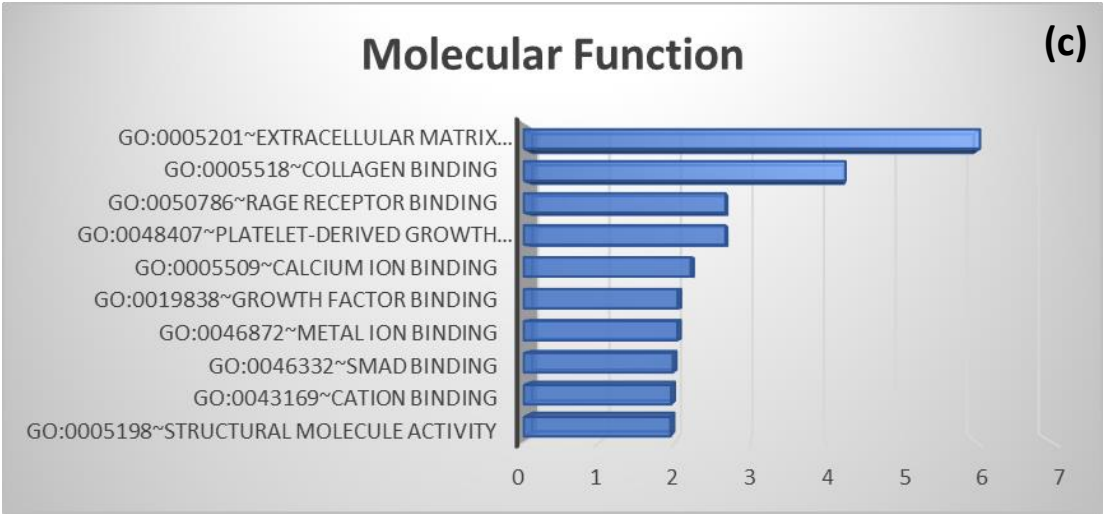
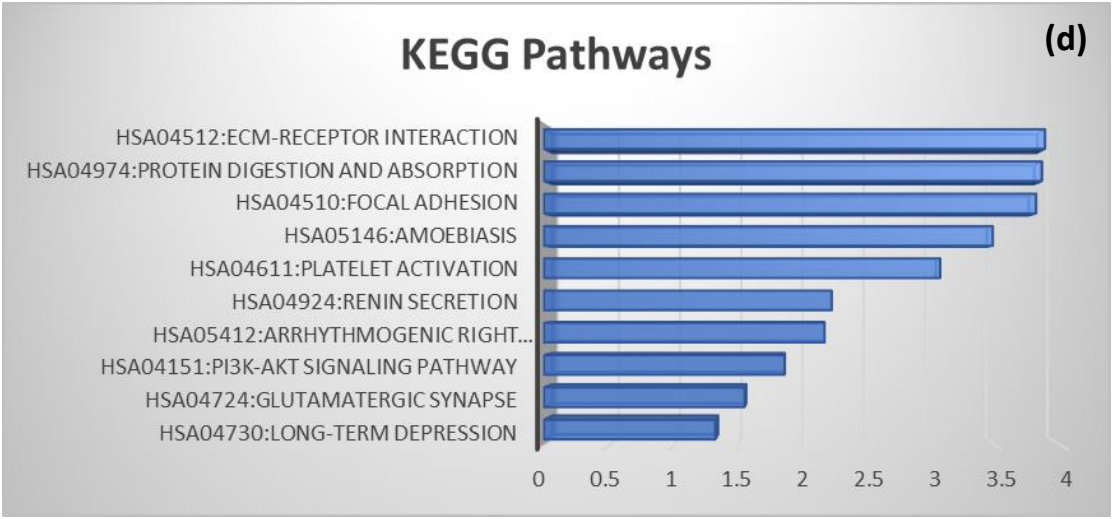

*Figure (2) The involvement degree of our 200 Top genes in different processes (by DAVID website)*

*(horizontal line:-log(p.value))* : (a)Biological process: Those processes that are vital for an organism to live and that shape its capacities for interacting with its environment. (b)Cellular component: The complex biomolecules and structures of cells and living organisms. (c): Molecular function: the elemental activities of a gene product at the molecular level, such as binding or catalysis. (d)KEGG Pathway: a collection of hand-drawn pathway maps that reflect our understanding of molecular interaction, reaction, and connection networks

*PPI network and the top 20 hub genes*

PPI information was acquired from the STRING database. Constructing and visualizing a PPI network of hub genes was conducted by Cytoscape software. Using the MCODE plugin in Cytoscape, we obtained the module with the highest score, which contained 54 nodes and 94 edges (figure 6). Then, the plugin cytoHubba was used to select the top 20 hub genes, which are in order of scores: COL1A1, COL1A2, COL3A1, ASPN, COL12A1, COL10A1, COL8A2, COL8A1, COL5A3, ITGA11, IGF1R, PLCB1, ERBB2, SMAD7, ITGB5, SPARC, MFAP5, PLCB4, S100A8, OMD

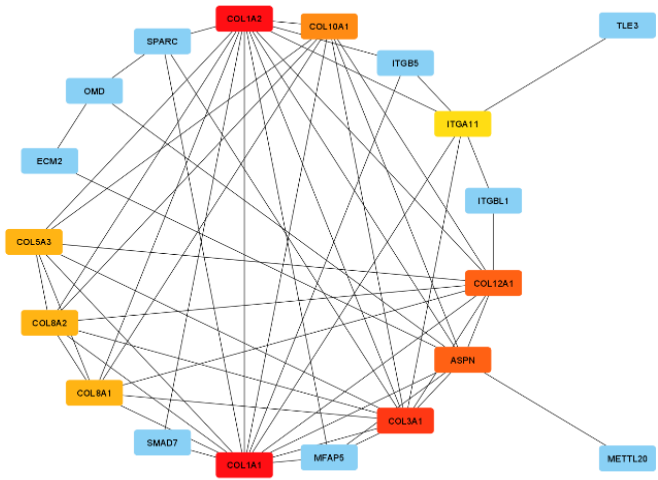

*Figure (3): Protein-protein interactions based on 200 Top genes*

*Table (1): Most related pathways to the 20 hub genes*

| Pathway | Percentage | FDR | Genes |
|---|---|---|---|
| Focal adhesion | 40 | 9.39E-07 | COL1A1, COL3A1, COL1A2, ITGB5, COL5A3, ERBB2, ITGA11, IGF1R |
| ECM-receptor interaction | 30 | 5.19E-06 | COL1A1, COL3A1, COL1A2, ITGB5, COL5A3, ITGA11 |
| Protein digestion and absorption | 30 | 5.19E-06 | COL1A1, COL3A1, COL1A2, COL5A3, COL12A1, COL10A1 |
| Amoebiasis | 30 | 9.91E-06 | COL1A1, COL3A1, PLCB4, COL1A2, COL5A3, PLCB1 |
| Platelet activation | 30 | 2.19E-05 | COL1A1, COL3A1, PLCB4, COL1A2, COL5A3, PLCB1 |
| PI3K-Akt signalling pathway | 35 | 1.25E-04 | COL1A1, COL3A1, COL1A2, ITGB5, COL5A3, ITGA11, IGF1R |
| Long-term depression | 15 | 0.046014 | PLCB4, PLCB1, IGF1R |
| Pathways in cancer | 20 | 0.238652 | PLCB4, ERBB2, PLCB1, IGF1R |
| Calcium signaling pathway | 15 | 0.287069 | PLCB4, ERBB2, PLCB1 |
| Proteoglycans in cancer | 15 | 0.298754 | ITGB5, ERBB2, IGF1R |
| Rap1 signaling pathway | 15 | 0.298754 | PLCB4, PLCB1, IGF1R |

*Hub genes analysis*

We used Funrich (version 3.1.3) program for various analyzes and charts based on our 20 top genes, such as biological process, cellular component, molecular function, Site of expression, heatmap, and COSMIC. (figure 4)

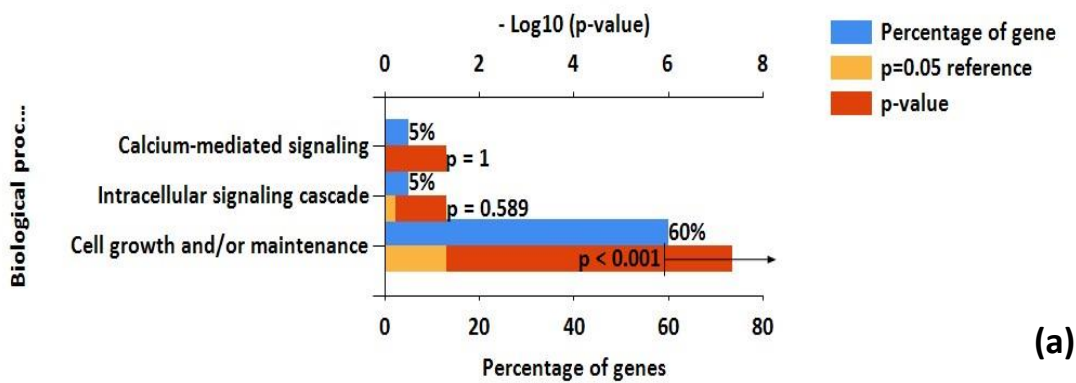

(a)

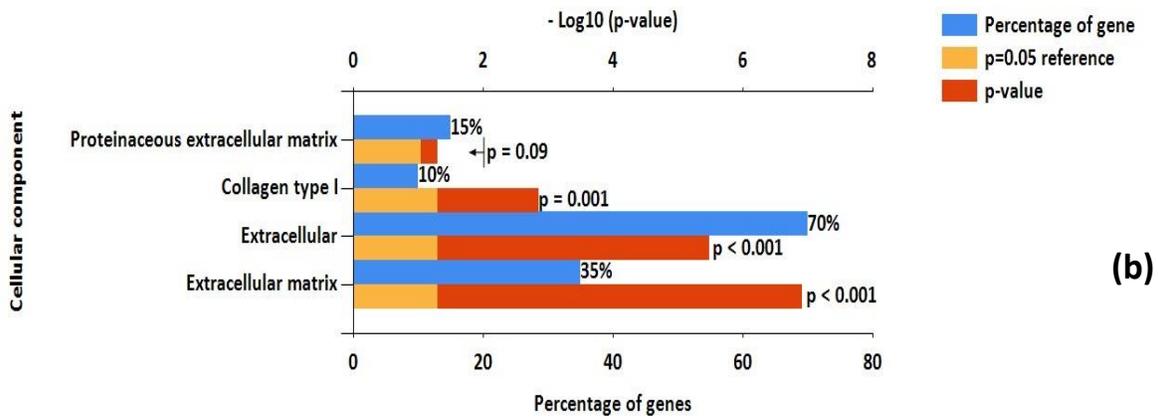

(b)

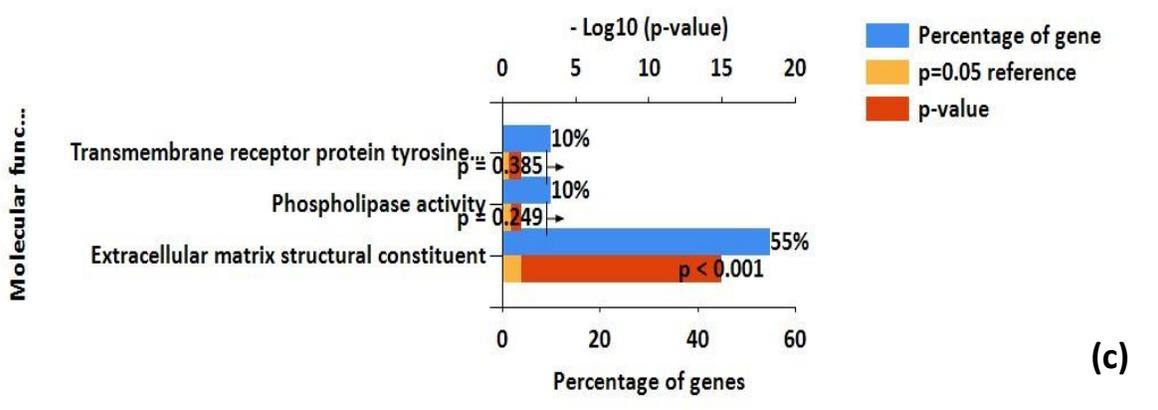

(c)

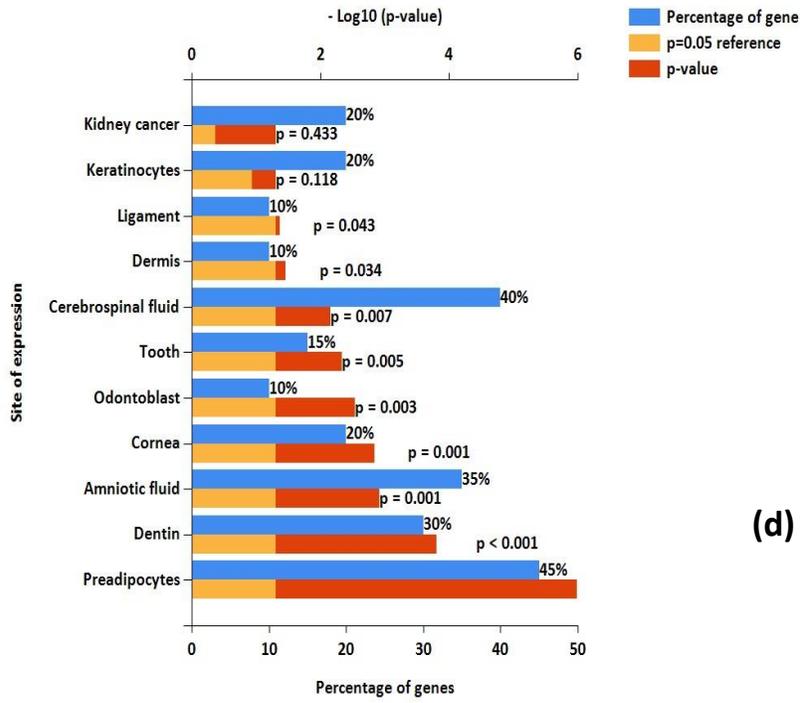

(d)

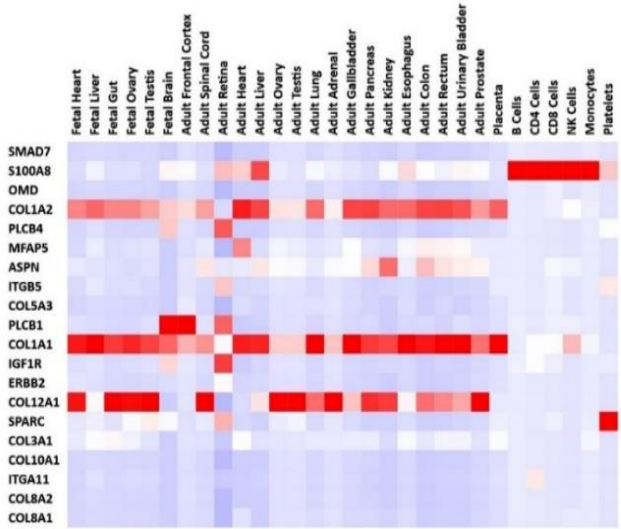

(e)

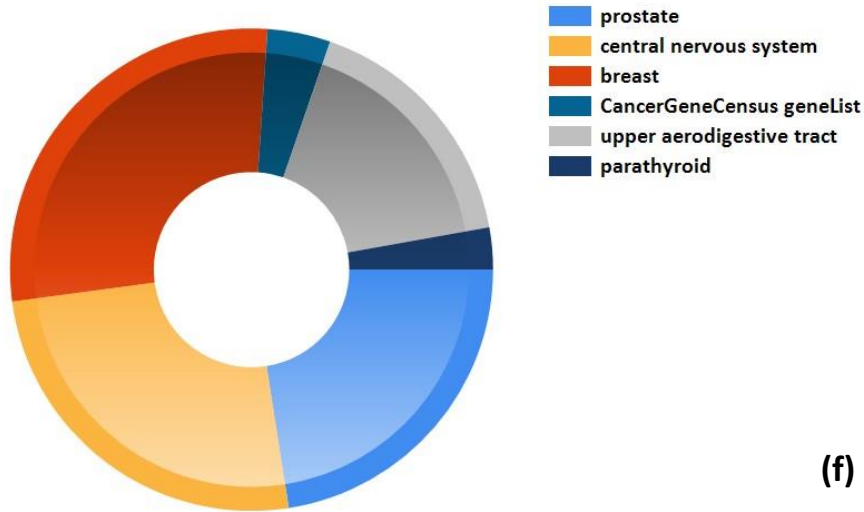

*Figure (4):* Analyzing 20 hub genes with Funrich software. (a): Biological process (b)Cellular component (c)Molecular function (d)Site of expression (normal tissues, cancer tissues, cell types and cell lines) (e)heat map gene expression (f) COSMIC: The Catalogue Of Somatic Mutations In Cancer, is the world's largest and most comprehensive resource for exploring the impact of somatic mutations in human cancer.

**Discussion**

Breast cancer is the most common cancer in women all over the world. Screening and diagnosing breast cancer at earlier stages are essential to improve patient survival and reduce treatment costs. However, the underlying mechanism regulating breast cancer aggressiveness remains poorly understood, and biomarkers for detecting early-stage breast cancer are still lacking.

In this study, we analyzed a microarray dataset from GEO, and Then the plugin cytoHubba was used to discover the top 20 hub genes. All these DEGs were classified into 11 protein categories, in which the top is Collagen. The functional enrichment analyses demonstrated that the DEGs were enriched in biological processes such as Collagen catabolic process, extracellular matrix organization, extracellular structure organization, and multicellular organism catabolic process. It has been demonstrated Collagen that has a significant role in the progression of CICs.

KEGG pathway analysis uncovered that the DEGs were significantly associated with ECM receptor interaction, Protein digestion, absorption, Focal adhesion, Amoebiasis, Protein digestion, absorption, and Amoebiasis and According to the obtained molecular function chart, the DEGs are associated with extracellular matrix structural constituent, collagen binding, platelet-derived growth factor binding, and RAGE receptor binding.

40 % of our hub genes, including COL1A1, COL3A1, COL1A2, ITGB5, COL5A3, ERBB2, ITGA11 and IGF1R, are involved Focal adhesion pathway and 30% of them that include COL1A1, COL3A1, COL1A2, ITGB5, COL5A3, ITGA11 are effective in ECM-receptor interaction pathway. By checking these two pathways on the KEGG website, we understand that they are entirely related.

The extracellular matrix (ECM) is a complex mixture of structural and functional macromolecules that plays a crucial role in tissue and organ morphogenesis and cell and tissue structure and function maintenance. Transmembrane molecules, primarily integrins, and proteoglycans, CD36, and other cell-surface-associated components mediate specific interactions between cells and the ECM. These interactions regulate cellular activities such as adhesion, migration, differentiation, proliferation, and apoptosis directly or indirectly. Integrins also serve as mechanoreceptors, forming a physical connection between the ECM and the cytoskeleton that transmits power. Integrins are glycosylated, heterodimeric transmembrane adhesion receptors with noncovalently bound alpha and beta subunits. ECM-receptor interactions are significant in the processes of tumour shedding, adhesion, degradation, movement, and hyperplasia, according to a 2019 study by Yulong Bao. It has been established that ECM plays a role in other cancers.

The extracellular matrix (ECM) is upregulated in prostate cancer tissue and plays a role in tumour invasion and metastasis in gastric cancer. In colorectal cancer cells, the ECM can promote epithelial-mesenchymal transition (EMT). The most lethal adult brain tumour is glioblastoma. It has pathological characteristics such as abnormal neovascularization and tumour cell infiltration. In this progression, the interactions between the ECM and the glioblastoma microenvironment are crucial. Tumour cells move through the ECM during metastasis, and the tumour suppressor nischarin can prevent cancer cell migration by interacting with a variety of proteins. According to some research, nischarin can prevent breast cancer cells from migrating and invading by altering the expression patterns of main adhesive proteins. Nischarin expression can reduce cells' ability to bind to the ECM, resulting in a reduction in invadopodium-mediated matrix degradation. Malignant tumours have a biological function called invasive metastasis. The high levels of expression of ECM proteins or genes in breast tumour tissues can lead to new cancer treatment ideas. (Bao, Wang et al.2019)

Cell-matrix adhesions are essential in various biological processes, including cell motility, proliferation, differentiation, gene regulation, and cell survival. Specialized structures known as focal adhesions form at cell-extracellular matrix contact points, A multi-molecular complex of junctional plaque proteins anchors bundles of actin filaments to transmembrane receptors of the integrin family. Some focal adhesion constituents are structural links between membrane receptors and the actin cytoskeleton, whereas others are signalling molecules such as protein kinases and phosphatases, their substrates, and various adapter proteins. Integrin signalling relies on the FAK (Focal Adhesion Kinase) and src proteins' non-receptor tyrosine kinase activities, as well as their adaptor protein functions, to trigger downstream signalling events. These signaling events reorganize the actin cytoskeleton, which is needed for changes in cell structure, motility, and gene expression. Growth factor binding to their respective receptors results in similar morphological changes and gene expression regulation, demonstrating the important crosstalk between adhesion and growth factor signaling. (genome website)

According to Damiano Cosimo Rigiracciolo's review from 2021, FAK circuitry has been identified as one of the vital transduction pathways leading to breast cancer progression based on the characterization of molecular alterations various forms of breast tumours. As a result, the evidence summarized in this analysis adds to the importance of FAK expression and activity in breast cancer proliferative and metastatic features. FAK is highly expressed in over 40% of breast tumours and is closely linked to resistance to chemotherapy and targeted therapies. The importance of FAK in promoting oncogenic pathways and inactivating tumour suppressive signals in mammary tumour progression and metastasis has also been identified. The importance of FAK in the functional interactions between breast cancer cells and the surrounding TME (tumour microenvironment) has recently been identified, leading to the acquisition of a high invasive phenotype. ACCORDING TO THIS DATA, high FAK expression and activity could be a promising biomarker for more detailed therapeutic strategies in breast cancer.

Furthermore, our findings demonstrate the significance of platelet activation and the PI3K-Akt signalling pathway in breast cancer. Platelets play an essential and beneficial function in primary hemostasis when the vessel wall's integrity is disrupted. Adhesion to adhesive macromolecules, such as Collagen and von Willebrand factor, or soluble platelet agonists, such as ADP, thrombin, and thromboxane A2, initiate platelet adhesion and activation at sites of vascular wall injury. Different agonists stimulate different receptors, resulting in increased intracellular Ca2+ concentrations, which stimulate platelet to shapeshift and granule secretion and eventually induce the "inside-out" signalling mechanism to activate the ligand-binding role of integrin alpha IIb beta 3. Platelet adhesion and aggregation are mediated by alpha IIb beta 3 binding to its ligands, primarily fibrinogen, and "outside-in" signalling is triggered, resulting in platelet spreading, additional granule secretion, stabilization of platelet adhesion and aggregation, and clot retraction. (genome website)

In a study by Chris E. Holmes, they have started looking into the unusual platelet phenotype caused by breast cancer and its treatment. Cancer cells' interactions with the primary tumour microenvironment are essential determinants of cancer progression to metastasis, and our findings indicate that platelets from cancer patients can influence the tumour microenvironment differently than platelets from healthy controls. Changes in platelet biology often cause differential responses to antiplatelet therapies such as ADP receptor antagonists. A better understanding of breast cancer and breast cancer treatment-related changes in platelet phenotype may significantly impact antiplatelet strategies to slow cancer progression and enable platelet inhibition therapy to be used safely in patients with heart disease and breast cancer. (Holmes, Levis et al. 2016)

Many types of cellular stimuli or toxic insults activate the phosphatidylinositol 3' kinase (PI3K)-Akt signalling pathway, which regulates fundamental cellular functions including transcription, translation, proliferation, development, and survival. Development factors induce class Ia and Ib PI3K isoforms as they bind to their receptor tyrosine kinase (RTK) or G protein-coupled receptors (GPCR). At the cell membrane, PI3K catalyzes the synthesis of phosphatidylinositol-3,4,5-triphosphate (PIP3). PIP3 then acts as a second messenger, assisting in the activation of Akt. By phosphorylating substrates involved in apoptosis, protein synthesis, metabolism, and the cell cycle, Akt can regulate critical cellular processes once activated.

According to Sherry X. Yang's study, the PI3K/AKT pathway is crucial in breast cancer pathogenesis, which has been discovered after decades of research; however, evidence on its effects on patient outcomes has been mixed. As previously stated, the translation is complicated by a heterogeneous breast cancer population whose tumours may vary in the magnitude of PI3K/AKT signalling dependence or have undergone different treatments and specific and changing treatment modalities that may have altered this pathway function. As a result, the findings for PIK3CA mutations and AKT activation in early-stage breast cancer prognosis were inconsistent. In the ER+ population, these changes were often linked to a positive or negligible outcome, while in the ER_ population, they were linked to a worse outcome. The effects of the PI3K/AKT pathway signalling in ER+ early breast cancer may have been overridden mainly by adjuvant endocrine therapy, which may have overridden the effects of the PI3K/AKT pathway signalling.

The Cellular component chart based on DAVID software shows that the DEGs are enriched in the intracellular parts. Many signals and intracellular proteins have been linked to developing various forms of CICs, such as Entosis (Mackay and Muller 2019). Entosis is initiated by the cell that will eventually be internalized (Overholtzer, Mailleux et al. 2007). This mechanism is also known as in-cell invasion, and it almost always leads to the death of the internal cell. Under standard tissue culture conditions, susceptible cell lines exhibit a low degree of entosis, but when cells are grown in matrix-detached conditions, higher rates of entosis are observed (Ishikawa, Ushida et al. 2015, Gol, Pena et al. 2018). Matrix detachment occurs before mitosis or apoptosis in standard culture conditions (Wang, He et al. 2016, Durgan, Tseng et al. 2017).
The discovery that several tumour cells could reliably engulf normal or non-transformed epithelial cells when admixed led to the hypothesis that entosis could act as a cell competition mechanism. (Mackay and Muller 2019)

CIC's roles in human cancers have been controversial (Huang, Chen et al. 2015); while early studies suggested a tumour-suppressing function based on its nature of cell death, later studies found that CIC-mediated engulfment can also promote tumour growth. The principle of cell rivalry recently resolved this disparity (Kroemer and Perfettini 2014, Sun, Luo et al. 2014). Heterogeneous tumours are characterized by the presence of numerous clones that compete for limited space and nutrients. CIC death slowed tumour development in the early stages. The winner tumour cell clones with oncogenic mutations were repeatedly internalized and outcompeted by CIC-mediated engulfment. Those that were less malignant, causing tumour growth to slow. According to CIC-induced aneuploidy, the winner cells have more opportunities to develop new mutations and malignant phenotypes, such as metastasis. As a result, during the late stages of cancer, the malignant winner clones with oncogenic mutations populate the tumour tissues and undergo distant metastasis (Sun, Huang et al. 2015)

COL1A1 (Collagen Type I Alpha 1 Chain), COL1A2 (Collagen Type I Alpha 2 Chain), COL3A1 (Collagen Type III Alpha 1 Chain), COL12A1 (Collagen Type XII Alpha 1 Chain), COL10A1 (Collagen Type X Alpha 1 Chain), COL8A2 (Collagen Type VIII Alpha 2 Chain), COL8A1 (Collagen Type VIII Alpha 1 Chain) and COL5A3 (Collagen Type V Alpha 3 Chain) were all in our ten hub genes so this shows the relationship between CICs and Collagen. Signal transduction pathways regulate cellular activity. External signals are received by cells via receptors and transmitted via a cascade, converting extracellular signals into intracellular signals and triggering physiological cellular reactions that control biological activities. Collagen, which is part of the ECM, affects cancer cell activity. Cancer cells reshape Collagen in the opposite direction, forming a self-reinforcing cell-collagen loop that promotes cancer progression. (Xu, Xu et al. 2019)

The ECM undergoes structural changes as a result of cancer incidence and progression. Collagen content and distribution are altered in cancer cells to better coordinate biological properties such as gene mutations, transcription factors, signalling pathways, and receptors. Cancer cell behaviour is influenced by the heterogeneity of mutated genes, one of the significant promoters of cancer cell behaviour. Oncogene mutations, classified as tumour suppressor genes or proto-oncogenes, affect the collagen conditions in the tumour matrix.

Cancer development is influenced by both cells and molecules in the tumour microenvironment. Collagen's role in cancer is a two-edged sword. Collagen, cancer cells and other matrix molecules, on the one side, form an interlocking loop that reinforces each other. (Xu, Xu et al. 2019)

ASPN is the fourth gene on the list of hub genes. ASPN is a small leucine-rich proteoglycan that belongs to the canonical class I family of extracellular matrix (ECM) proteins with a special Drepeat in its Nterminal region (Sasaki, Takagane et al. 2021) ASPN is mainly located in the ECM of cartilage that surrounds skeletal tissue (Jiang, Liu et al. 2019). ASPN has been studied extensively in osteoarthritis and is tightly regulated in cartilage and has been linked to breast ductal carcinoma local invasion (Castellana, Escuin et al. 2012). Furthermore, ASPN is overexpressed in pancreatic ductal adenocarcinoma, indicating that it may be helpful for diagnostic and therapeutic purposes (Turtoi, Musmeci et al. 2011). ASPN also affects epidermal growth factor receptor (EGFR) receptor signalling, which helps GC cells expand and migrate (Ding, Zhang et al. 2015).

ITGA11 (Integrin Subunit Alpha 11) and ITGB5 are two integrin subunits on our list (Integrin Subunit Beta 5).
Understanding new developments in the transcriptome and pathways can lead to the discovery of new cancer biomarkers. Integrin 11 (ITGA11), a member of the integrin family, is involved in various biological processes, including metastasis, embryogenesis, hemostasis, immune response, tissue repair, cancer development, tumour angiogenesis, and therapy resistance (Bos, Van der Groep et al. 2003, Wong, Gilkes et al. 2011). Integrin mutations disrupt cancer cell adhesion and extracellular matrix assembly, potentially resulting in tumour metastasis (Schindl, Schoppmann et al. 2002). Tyrosine kinase receptors facilitate cancer cell proliferation and differentiation, and integrins interact with them.

Integrins are heterodimeric cell surface adhesion receptors that are made up of two subunits: and. A combinatorial association of 18 and 8 subunits results in 24 distinct integrin heterodimers (Van der Flier and Sonnenberg 2001). ECM ligands can bind to the subunit and activate intracellular signalling events via the subunit, allowing extracellular and intracellular events to be integrated for cell motility and invasion. Several integrins are expressed at low or undetectable levels in adult epithelia, but they are upregulated in tumours (Ju, Godet et al.). Integrin 11 (ITGA11) is expressed in various tissues in the fetus, but it disappears with maturation in adult tissues (Hynes 2002).

The Site of expression chart shows that these 20 genes are associated with preadipocytes, Dentin, Amniotic fluid, cornea, odontoblast, tooth, cerebrospinal fluid, Dermis, ligament, keratinocytes, and kidney cancer.

Also, we understood that these genes are related to breast, prostate, central nervous system, upper aerodigestive tract, Cancer Gene Census gene list, and parathyroid from the COSMIC chart.

In conclusion, according to the results from analyzing the 20 hub genes, we understood that overexpression of our Top genes is effective in focal adhesion, ECM-receptor interaction, platelet activation and PI3K-Akt signalling pathway, which shows that changes in these pathways could be the reason for the overexpression of CICs in breast cancer. Involvement of our most scored genes in these pathways could indicate their effect on increasing the incidence of CICs and further its impact on tumour cell malignancy in breast cancer. The high expression levels of ECM proteins or genes in breast tumour tissues can lead to new cancer treatment ideas. We believe that these genes and pathways may be possible breast cancer markers, but further research is needed to confirm the mechanisms of tumorigenesis and growth. Also, from the hub genes, there is a strong connection between breast cancer and Collagen. By inducing cancer cell proliferation, migration, and metastasis. In conclusion, the relationship between Collagen and cancer is only partially understood, and further research is required to elucidate detailed collagen biological mechanisms in cancer tissue that can precisely control collagen balance for entire treatment gain. When combined with other treatment options, this new approach can help patients live longer and have a better quality of life.

**چکیده**

سرطان پستان شایع ترین نوع سرطان بین زنان سراسر جهان است. کشف نشانگرهای زیستی این بیماری بسیار مهم قلمداد می شود و ساختار سلول در سلول (CICs) می تواند یکی از آنها باشد که ممکن است بتواند به عنوان یک نشانگر بدخیمی تومور مورد استفاده قرار گیرد. CICs از این نظر غیر معمول است که سلولهای سالم از نظر مورفولوژی را در سلول دیگری نگه می دارد. آنها در سرطان های مختلف انسان یافت می شوند و در نتیجه فعل و انفعال سلول-سلول فعال هستند و انواع مختلفی دارد. در این مطالعه، ما داده های ریز آرایه (GSE103865) GEO را برای ارزیابی ژنتیکی میزان CICs در نمونه های بدست آمده از بیماران مبتلا به سرطان پستان برای درک رابطه بین میزان CICs و پیش آگهی این سرطان، آنالیز کردیم. پیش پردازش با استفاده از نرم افزار R انجام شد. از وبسایت DAVID برای آنالیز هستی شناسی ژن (GO) و مسیر ژن و ژنوم (KEGG) استفاده شد. فعل و انفعالات پروتئین-پروتئین (PPIs) از DEG های بدست آمده با استفاده از وبسایت STRING بررسی شد و در Cytoscape و cytoHubba غربال شدند. نتایج حاصل از آنالیز ۲۰ hubgenes نشان دهنده آن است که افزایش بیان ژنهای برتر در مسیرهای چسبندگی کانونی، تعامل گیرنده های ECM(Extra Cellular Matrix)، فعال سازی پلاکت ها و مسیرP13K-Akt موثر است. تحلیل این داده ها به همراه تحقیقات انجام شده دیگر، ژن های مختلفی را در تشکیل CICs شناسایی کرده است و باعث ایجاد ایده در مورد علت تشکیل آنها و چگونگی کمک به سرطان پستان می کند.